\documentclass[aps,pra,twocolumn,superscriptaddress,reprint,10pt]{revtex4-1}
\usepackage{amsmath}
\usepackage{amsthm}
\usepackage{amssymb}
\usepackage{hyperref}
\newtheorem{thm}{Theorem}[section]

\newtheorem{lem}[thm]{Lemma}

\theoremstyle{definition}
\newtheorem{defn}[thm]{Definition}
\theoremstyle{remark}
\newtheorem{rem}[thm]{Remark}
\numberwithin{equation}{section}

\begin{document}
\title{Passive states as optimal inputs for single-jump lossy quantum channels}
\author{Giacomo De Palma}
\affiliation{NEST, Scuola Normale Superiore and Istituto Nanoscienze-CNR, I-56127 Pisa,
Italy.}
\affiliation{INFN, Pisa, Italy}
\author{Andrea Mari}
\affiliation{NEST, Scuola Normale Superiore and Istituto Nanoscienze-CNR, I-56127 Pisa,
Italy.}
\author{Seth Lloyd}
\affiliation{Research Laboratory of Electronics, Massachusetts Institute of Technology, Cambridge, MA 02139, USA.}
\affiliation{Department of Mechanical Engineering, Massachusetts Institute of Technology, Cambridge, MA 02139, USA.}
\author{Vittorio Giovannetti}
\affiliation{NEST, Scuola Normale Superiore and Istituto Nanoscienze-CNR, I-56127 Pisa,
Italy.}

\begin{abstract}
The passive states of a quantum system minimize the average energy among all the states with a given spectrum.
We prove that passive states are the optimal inputs of single-jump lossy quantum channels.
These channels arise from a weak interaction of the quantum system of interest with a large Markovian bath in its ground state, such that the interaction Hamiltonian couples only consecutive energy eigenstates of the system.
We prove that the output generated by any input state $\rho$ majorizes the output generated by the passive input state $\rho_0$ with the same spectrum of $\rho$.
Then, the output generated by $\rho$ can be obtained applying a random unitary operation to the output generated by $\rho_0$.
This is an extension of De Palma et al., IEEE Trans. Inf. Theory 62, 2895 (2016), where the same result is proved for one-mode bosonic Gaussian channels.
We also prove that for finite temperature this optimality property can fail already in a two-level system, where the best input is a coherent superposition of the two energy eigenstates.
\end{abstract}
\maketitle

\section{Introduction}
The passive states \cite{pusz1978passive,lenard1978thermodynamical} of a quantum system are the states diagonal in the eigenbasis of the Hamiltonian, with eigenvalues decreasing as the energy increases.
They minimize the average energy among all the states with a given spectrum, and hence no work can be extracted from them on average with unitary operations \cite{janzing2006computational}. For this reason they play a key role in the recently emerging field of quantum thermodynamics (see \cite{vinjanampathy2015quantum,goold2015role} for a review).

Majorization \cite{marshall2010inequalities} is the order relation between quantum states induced by random unitary operations, i.e. a state $\hat{\sigma}$ is majorized by a state $\hat{\rho}$ iff $\hat{\sigma}$ can be obtained applying random unitaries to $\hat{\rho}$.
Majorization theory is ubiquitous in quantum information.
Its very definition suggests applications in quantum thermodynamics \cite{goold2015role,gour2015resource,horodecki2013fundamental}, where the goal is determining the set of final states that can be obtained from a given initial state with a given set of operations.
In the context of quantum entanglement, it also determines whether it is possible to convert a given bipartite pure state into another given pure state by means of local operations and classical communication \cite{nielsen1999conditions,nielsen2001majorization}.
More in the spirit of this paper, majorization has proven to be crucial in the longstanding problem of the determination of the classical communication capacity of quantum gauge-covariant bosonic Gaussian channels \cite{giovannetti2014ultimate}, and the consequent proof of the optimality of Gaussian states for the information encoding.
Indeed, a turning point has been the proof of a majorization property: the output of any of these channels generated by any input state is majorized by the output generated by the vacuum \cite{mari2014quantum,giovannetti2015majorization} (see also \cite{holevo2015gaussian} for a review).
This fundamental result has been extended and linked to the notion of passive states in Ref. \cite{de2015passive}, where it is proved that these states optimize the output of any one-mode quantum Gaussian channel, in the sense that the output generated by a passive state majorizes the output generated by any other state with the same spectrum. Moreover, the same channels preserve the majorization relation when applied to passive states \cite{jabbour2015majorization}.

In this paper we extend the result of Ref. \cite{de2015passive} to a large class of lossy quantum channels.
Lossy quantum channels arise from a weak interaction of the quantum system of interest with a large Markovian bath in its zero-temperature (i.e. ground) state.
We prove that passive states are the optimal inputs of these channels.
Indeed, we prove that the output $\Phi\left(\hat{\rho}\right)$ generated by any input state $\hat{\rho}$ majorizes the output $\Phi\left(\hat{\rho}^\downarrow\right)$ generated by the passive input state $\hat{\rho}^\downarrow$ with the same spectrum of $\hat{\rho}$.
Then, $\Phi\left(\hat{\rho}\right)$ can be obtained applying a random unitary operation to $\Phi\left(\hat{\rho}^\downarrow\right)$, and it is more noisy than $\Phi\left(\hat{\rho}^\downarrow\right)$.
Moreover, $\Phi\left(\hat{\rho}^\downarrow\right)$ is still passive, i.e. the channel maps passive states into passive states.

In the context of quantum thermodynamics, this result puts strong constraints on the possible spectrum of the output of lossy channels.
It can then be useful to determine which output states can be obtained from an input state with a given spectrum in a resource theory with the lossy channel among the allowed operations.
The Gaussian analogue of our result has been crucial for proving that Gaussian input states minimize the output entropy of the one-mode Gaussian quantum attenuator for fixed input entropy \cite{de2016gaussian}.
Our result can find applications in the proof of similar entropic inequalities on the output states of lossy channels in the same spirit of the quantum Entropy Power Inequalities of \cite{konig2014entropy,de2014generalization,de2015multimode,audenaert2015entropy}, and then determine their classical capacity.

Our result applies to all the interactions of a quantum system with a heat bath such that the reduced system dynamics can be modeled by a master equation \cite{schaller2014open,breuer2007theory} and the following hypotheses are satisfied:
\begin{enumerate}
\item The Hamiltonian of the system is non-degenerate.
\item The system-bath interaction Hamiltonian couples only consecutive eigenstates of the Hamiltonian of the system alone.
\item If the system starts in its maximally mixed state, its reduced state remains passive.
\item The bath starts in its ground (i.e. zero temperature) state.
\end{enumerate}
The first assumption is satisfied by a large class of quantum systems, and it is usually taken for granted in both quantum thermodynamics and quantum statistical mechanics \cite{gogolin2015equilibration}.
The second assumption is also satisfied by a large class of quantum systems.
The third assumption means that the interaction cannot generate population inversion if the system is initialized in the infinite-temperature state, as it is for most physical systems.
The fourth assumption is for example satisfied by the interaction of a quantum system with an optical bath at room temperature.
Indeed, $\hbar\omega\gg k_BT$ for $\omega$ in the optical range and $T\approx300^\circ K$, hence the state of the bath at room temperature is indistinguishable from the vacuum.

These assumptions turn out to be necessary.
Indeed, dropping any of them it is possible to find explicit  counterexamples for which passive inputs are not optimal choices for output majorization.

The manuscript is organized as follows. In Sec.~\ref{secmaj} we briefly recall some basics facts about majorization and the notion of passive states. The main result of the paper
 is instead presented in Sec.~\ref{secopt} where we first  define in a rigorous way the class of lossy maps we are interested in and then proceed with a formal proof the optimality for passive states.
Section~\ref{seccount} is instead devoted to counterexamples.
In particular in Sec.~\ref{seccatt} we show that for the two-mode bosonic Gaussian quantum-limited attenuator, whose associated Hamiltonian is degenerate, no majorization relations can be ascribed to the passive states.
In Sec.\ref{sec2step} instead a counterexample is provided for a two-qubit lossy map with two different choices of the Hamiltonian.
In Section \ref{+jumps} the Hamiltonian is non-degenerate, but the process involves quantum jumps of more than one energy step.
In Section \ref{degH} only quantum jumps of one energy step are allowed, but the Hamiltonian becomes degenerate.
In Sec.~\ref{secfin} we analyze the case of a map where the bath temperature is not zero.
We show that the optimal input states are a pure coherent superposition of the Hamiltonian eigenstates, hence non passive. Conclusions and comments are presented in Sec.~\ref{seccon} while technical derivations are presented in the appendices.

\section{Majorization}\label{secmaj}
Majorization is a concept that gives a precise meaning to the proposition ``the quantum state $\hat{\rho}$ (or the probability distribution $p$) is less disordered than the quantum state $\hat{\sigma}$ (or the probability distribution $q$)''.
The interested reader can find more details in the dedicated book \cite{marshall2010inequalities}.

Let us start with the definition for probability distributions:
\begin{defn}[Majorization]\label{majd}
Let $p$ and $q$ be two discrete probability distributions on a set of $d\in\mathbb{N}$ elements with
\begin{equation}
p_1\geq\ldots\geq p_d\;,\qquad q_1\geq\ldots\geq q_d\;.
\end{equation}
We say that $p$ majorizes $q$, or $p\succ q$, iff
\begin{equation}
\sum_{i=1}^n p_i\geq\sum_{i=1}^n q_i\quad\forall\;n=1,\ldots,d\;.
\end{equation}
\end{defn}
Definition \ref{majd} can be easily extended to quantum states:
\begin{defn}
Let $\hat{\rho}$ and $\hat{\sigma}$ be quantum states acting on $\mathbb{C}^d$ with eigenvalues $p_1\geq\ldots\geq p_d$ and $q_1\geq\ldots\geq q_d$, respectively.
We say that $\hat{\rho}$ majorizes $\hat{\sigma}$, or $\hat{\rho}\succ\hat{\sigma}$, iff $p\succ q$.
\end{defn}
From an operational point of view, and for the applications in quantum information and quantum thermodynamics, it is useful to express majorization as the order relation induced by random unitary operations (see Sec. II.C of \cite{wehrl1978general}):
\begin{thm}
Given two quantum states $\hat{\rho}$ and $\hat{\sigma}$, the following conditions are equivalent:
\begin{enumerate}
  \item $\hat{\rho}\succ\hat{\sigma}$;
  \item For any convex function $f:[0,1]\to\mathbb{R}$,
  \begin{equation}\label{Trf}
  \mathrm{Tr}\;f\left(\hat{\rho}\right)\geq\mathrm{Tr}\;f\left(\hat{\sigma}\right)\;.
  \end{equation}
  Taking $f(x)=x\ln x$ and $f(x)=x^p,\;p>1$, \eqref{Trf} implies that the von Neumann and all the R\'enyi entropies \cite{holevo2013quantum} of $\hat{\rho}$ are lower than the corresponding ones of $\hat{\sigma}$;
  \item $\hat{\sigma}$ can be obtained applying to $\hat{\rho}$ random unitary operators, i.e. there exist $n\in\mathbb{N}$, a probability distribution $p$ on $\left\{1,\ldots,n\right\}$ and a family of unitary operators $\left\{\hat{U}_1,\ldots,\hat{U}_n\right\}$ such that
  \begin{equation}\label{majru}
  \hat{\sigma}=\sum_{i=1}^np_i\;\hat{U}_i\;\hat{\rho}\;\hat{U}_i^\dag\;.
  \end{equation}
\end{enumerate}
\end{thm}

\subsection{Passive states}\label{rearrangement}
We consider a $d$-dimensional quantum system with non-degenerate Hamiltonian
\begin{equation}\label{Hpass}
\hat{H}=\sum_{i=1}^{d} E_i\;|i\rangle\langle i|\;,\qquad \langle i|j\rangle=\delta_{ij}\;,\qquad E_1<\ldots<E_{d}\;.
\end{equation}
A self-adjoint operator is \emph{passive} \cite{pusz1978passive,lenard1978thermodynamical} if it is diagonal in the eigenbasis of the Hamiltonian and its eigenvalues decrease as the energy increases.
\begin{defn}[Passive rearrangement]\label{passr}
Let $\hat{X}$ be a self-adjoint operator with eigenvalues $x_1\geq\ldots\geq x_d$.
We define its passive rearrangement as
\begin{equation}
\hat{X}^\downarrow:=\sum_{i=1}^d x_i\;|i\rangle\langle i|\;,
\end{equation}
where $\left\{|i\rangle\right\}_{i=1,\ldots,n}$ is the eigenbasis of the Hamiltonian \eqref{Hpass}.
Of course, $\hat{X}=\hat{X}^\downarrow$ for any passive operator.
\end{defn}
\begin{rem}\label{passiveU}
The passive rearrangement is unitarily invariant, i.e.
\begin{equation}
\left(\hat{U}\;\hat{X}\;\hat{U}^\dag\right)^\downarrow=\hat{X}^\downarrow
\end{equation}
for any self-adjoint operator $\hat{X}$ and any unitary operator $\hat{U}$.
\end{rem}
\begin{rem}
The passive rearrangement of any rank-$n$ projector $\hat{\Pi}_n$ is the projector onto the first $n$ energy eigenstates:
\begin{equation}\label{Pin*}
\hat{\Pi}_n^\downarrow=\sum_{i=1}^n|i\rangle\langle i|\;.
\end{equation}
\end{rem}
\begin{rem}
It is easy to show that passive quantum states minimize the average energy among all the states with a given spectrum, i.e.
\begin{equation}
\mathrm{Tr}\left[\hat{H}\;\hat{U}\;\hat{\rho}\;\hat{U}^\dag\right]\geq\mathrm{Tr}\left[\hat{H}\;\hat{\rho}^\downarrow\right]\qquad\forall\;\hat{U}\;\text{unitary}\;.
\end{equation}
\end{rem}

\section{Optimality of passive states for lossy channels} \label{secopt}
The most general master equation that induces a completely positive Markovian dynamics is \cite{breuer2007theory,schaller2014open}
\begin{equation}
\frac{d}{dt}\hat{\rho}(t)=\mathcal{L}\left(\hat{\rho}(t)\right)\;,
\end{equation}
where the generator $\mathcal{L}$ has the Lindblad form
\begin{equation}\label{Ldef}
\mathcal{L}\left(\hat{\rho}\right)=-i\left[\hat{H}_{LS},\;\hat{\rho}\right]+\sum_{\alpha=1}^{\alpha_0}\left(\hat{L}_\alpha\;\hat{\rho}\;\hat{L}_\alpha^\dag-\frac{1}{2}\left\{\hat{L}_\alpha^\dag\hat{L}_\alpha,\;\hat{\rho}\right\}\right)\;,
\end{equation}
where $\alpha_0\in\mathbb{N}$.
This dynamics arises from a weak interaction with a large Markovian bath in the rotating-wave approximation \cite{breuer2007theory,schaller2014open}.
In this case, $\hat{H}_{LS}$ commutes with the Hamiltonian $\hat{H}$, i.e. $\hat{H}_{LS}$ only shifts the energies of $\hat{H}$:
\begin{equation}\label{HLS}
\hat{H}_{LS}=\sum_{i=1}^d \delta E_i\;|i\rangle\langle i|\;.
\end{equation}
As anticipated in the introduction, we suppose that the bath starts in its ground state and that the interaction Hamiltonian $\hat{V}_{SB}$ couples only neighbouring energy levels of the system:
\begin{equation}
\hat{V}_{SB}=\sum_{i=1}^d|i\rangle_S\langle i|\otimes \hat{V}^B_i+\sum_{i=1}^{d-1}\left(|i\rangle_S\langle i+1|\otimes \hat{W}^B_i+\text{h.c.}\right)\;.
\end{equation}
Here the $\hat{V}_i^B$ are generic self-adjoint operators, while the $\hat{W}_i^B$ are completely generic operators.
In the rotating-wave approximation only the transitions that conserve the energy associated to the noninteracting Hamiltonian are allowed.
If the bath is in its ground state, it cannot transfer energy to the system, and only the transitions that decrease its energy are possible.
Then, each Lindblad operator $\hat{L}_\alpha$ can induce either dephasing in the energy eigenbasis:
\begin{equation}\label{dephase}
\hat{L}_\alpha=\sum_{i=1}^d a_i^\alpha\;|i\rangle\langle i|\;,\qquad a_i^\alpha\in\mathbb{C}\;,\qquad\alpha=1,\ldots,\,\alpha_0\;,
\end{equation}
or decay toward the ground state with quantum jumps of one energy level:
\begin{equation}\label{jump}
\hat{L}_\alpha=\sum_{i=1}^{d-1} b_i^\alpha\;|i\rangle\langle i+1|\;,\qquad b_i^\alpha\in\mathbb{C}\;,\qquad\alpha=1,\ldots,\,\alpha_0\;.
\end{equation}

It is easy to show that, if $\hat{\rho}$ is diagonal in the energy eigenbasis, also $\mathcal{L}\left(\hat{\rho}\right)$ is diagonal in the same basis, hence $e^{t\mathcal{L}}\left(\hat{\rho}\right)$ remains diagonal for any $t$.

As anticipated in the introduction, we also suppose that the quantum channel $e^{t\mathcal{L}}\left(\hat{\rho}\right)$ sends the maximally mixed state into a passive state.
As a consequence, the generator $\mathcal{L}$ maps the identity into a passive operator (see Section \ref{passt}).

To see explicitly how this last condition translates on the coefficients $b_i^\alpha$, we compute
\begin{equation}
\mathcal{L}\left(\hat{\mathbb{I}}\right)=\sum_{i=1}^d\left(\sum_\alpha\left(\left|b_i^\alpha\right|^2-\left|b_{i-1}^\alpha\right|^2\right)\right)|i\rangle\langle i|\;,
\end{equation}
where for simplicity we have set $b_0^\alpha=b_d^\alpha=0$, and the operator is passive iff the function
\begin{equation}\label{r_i}
r_i:=\sum_\alpha\left|b_i^\alpha\right|^2\;,\qquad i=0,\,\ldots,\,d
\end{equation}
is concave in $i$.

The main result of this paper is that passive states optimize the output of the quantum channel generated by any dissipator of the form \eqref{Ldef} satisfying \eqref{HLS} and with Lindblad operators of the form \eqref{dephase} or \eqref{jump} such that the function \eqref{r_i} is concave.
We will prove that the output $e^{t\mathcal{L}}\left(\hat{\rho}\right)$ generated by any input state $\hat{\rho}$ majorizes the output $e^{t\mathcal{L}}\left(\hat{\rho}^\downarrow\right)$ generated by the passive state $\hat{\rho}^\downarrow$ with the same spectrum of $\hat{\rho}$, i.e. for any $t\geq0$
\begin{equation}\label{optimaldef}
e^{t\mathcal{L}}\left(\hat{\rho}\right)\prec e^{t\mathcal{L}}\left(\hat{\rho}^\downarrow\right)\;.
\end{equation}
We can substitute $\hat{\rho}\mapsto\hat{U}\,\hat{\rho}\,\hat{U}^\dag$ into \eqref{optimaldef}, with $\hat{U}$ a unitary operator.
We get that for any quantum state $\hat{\rho}$ and any unitary operator $\hat{U}$
\begin{equation}
e^{t\mathcal{L}}\left(\hat{U}\;\hat{\rho}\;\hat{U}^\dag\right)\prec e^{t\mathcal{L}}\left(\hat{\rho}^\downarrow\right)\;,
\end{equation}
where we have used Remark \ref{passiveU}.
Moreover, for any $t\geq0$ the state $e^{t\mathcal{L}}\left(\hat{\rho}^\downarrow\right)$ is still passive, i.e. the quantum channel $e^{t\mathcal{L}}$ preserves the set of passive states.
The proof closely follows \cite{de2015passive}, and is contained in the next section.

\subsection{Proof of the main result}\label{mainproof}
Let us define
\begin{equation}
\hat{\rho}(t)=e^{t\mathcal{L}}\left(\hat{\rho}\right)\;.
\end{equation}
The quantum states with non-degenerate spectrum are dense in the set of all quantum states.
Besides, the spectrum is a continuous function of the operator, and any linear map is continuous.
Then, without loss of generality we can suppose that $\hat{\rho}$ has non-degenerate spectrum.
Let $p_1(t)\geq\ldots\geq p_{d}(t)$ be the eigenvalues of $\hat{\rho}(t)$, and let
\begin{equation}
s_n(t)=\sum_{i=1}^n p_i(t)\;,\qquad n=1,\ldots,\,d\;.
\end{equation}
Let instead
\begin{equation}
p_i^\downarrow(t)=\langle i|e^{t\mathcal{L}}\left(\hat{\rho}^\downarrow\right)|i\rangle\;,\qquad i=1,\,\ldots,\,d
\end{equation}
be the eigenvalues of $e^{t\mathcal{L}}\left(\hat{\rho}^\downarrow\right)$, and
\begin{equation}
s_n^\downarrow(t)=\sum_{i=1}^n p_i^\downarrow(t)\;,\qquad n=1,\,\ldots,\,d\;.
\end{equation}
Notice that $p(0)=p^\downarrow(0)$ and then $s(0)=s^\downarrow(0)$, where
\begin{eqnarray}
p(t) &=& \left(p_1(t),\ldots,p_d(t)\right)\;,\\
s(t) &=& \left(s_1(t),\ldots,s_d(t)\right)\;.
\end{eqnarray}
The proof comes from:
\begin{lem}\label{deg}
The spectrum of $\hat{\rho}(t)$ can be degenerate at most in isolated points.
\begin{proof}
See Section \ref{proofdeg} in the Appendix.
\end{proof}
\end{lem}
\begin{lem}\label{lemma1}
$s(t)$ is continuous in $t$, and for any $t\geq0$ such that $\hat{\rho}(t)$ has non-degenerate spectrum it satisfies
\begin{equation}\label{sdot}
\frac{d}{dt}s_n(t)\leq\lambda_n(s_{n+1}(t)-s_n(t))\;,\qquad n=1,\,\ldots,\,d-1\;,
\end{equation}
where
\begin{equation}
\lambda_n=\mathrm{Tr}\left[\hat{\Pi}_n^\downarrow\;\mathcal{L}\left(\hat{\mathbb{I}}\right)\right]\geq0\;.
\end{equation}
\begin{proof}
See section \ref{prooflemma1} in the Appendix.
\end{proof}
\end{lem}
\begin{lem}\label{lemma2}
If $s(t)$ is continuous in $t$ and satisfies \eqref{sdot}, then $s_n(t)\leq s_n^\downarrow(t)$ for any $t\geq0$ and $n=1,\,\ldots,\,d$.
\begin{proof}
See section \ref{prooflemma2} in the Appendix.
\end{proof}
\end{lem}
Lemma \ref{lemma2} implies that for any $t\geq0$ the quantum channel $e^{t\mathcal{L}}$ preserves the set of passive states.
Indeed, let us choose the initial state $\hat{\rho}$ already passive.
Then, $s_n(t)$ is the sum of the $n$ largest eigenvalues of $e^{t\mathcal{L}}\left(\hat{\rho}\right)$.
Recalling that $e^{t\mathcal{L}}\left(\hat{\rho}\right)$ is diagonal in the Hamiltonian eigenbasis, $s_n^\downarrow(t)$ is the sum of the eigenvalues corresponding to the first $n$ eigenstates of the Hamiltonian $|1\rangle,\;\ldots,\;|n\rangle$, so that $s_n^\downarrow(t)\leq s_n(t)$.
However, Lemma \ref{lemma2} implies $s_n(t)=s_n^\downarrow(t)$ for $n=1,\,\ldots,\,d$, then $p_n(t)=p_n^\downarrow(t)$ and $e^{t\mathcal{L}}\left(\hat{\rho}\right)$ preserves the set of passive states for any $t$.

Then, for the definition of majorization and Lemma \ref{lemma2} again,
\begin{equation}
e^{t\mathcal{L}}\left(\hat{\rho}\right)\prec e^{t\mathcal{L}}\left(\hat{\rho}^\downarrow\right)
\end{equation}
for any $\hat{\rho}$, and the passive states are the optimal inputs for the channel.

\section{Counterexamples} \label{seccount}
The maximally mixed state is passive.
Then, if we want the channel to preserve the set of passive states, it must send the maximally mixed state into a passive state, and this hypothesis is necessary.
Also the hypotheses of non-degenerate Hamiltonian, quantum jumps of only one energy step and zero temperature are necessary.
Indeed, for each of them we present a counterexample violating only that hypothesis and for which Eq. \eqref{optimaldef} does not hold.

\subsection{Gaussian attenuator with degenerate Hamiltonian}
\label{seccatt}
The hypothesis of non-degenerate Hamiltonian is necessary for the optimality of passive states.
Indeed, in this section we provide an explicit counterexample with degenerate Hamiltonian: the two-mode bosonic Gaussian quantum-limited attenuator \cite{ferraro2005gaussian,holevo2013quantum}.

Let us fix $N\geq5$, and let $\mathcal{H}_N$ be the span of the first $N+1$ Fock states $\left\{|0\rangle,\ldots,|N\rangle\right\}$ of the Hilbert space of the harmonic oscillator.
Let us consider the restriction to $\mathcal{H}_N$ of the Hamiltonian of the harmonic oscillator
\begin{equation}
\hat{H}=\sum_{i=1}^N i\;|i\rangle\langle i|\;,\qquad\langle i|j\rangle=\delta_{ij}\;,
\end{equation}
and the Lindbladian
\begin{equation}\label{Latt}
\mathcal{L}\left(\hat{\rho}\right)=\hat{a}\;\hat{\rho}\;\hat{a}^\dag-\frac{1}{2}\left\{\hat{a}^\dag\hat{a},\;\hat{\rho}\right\}\;,
\end{equation}
where $\hat{a}$ is the ladder operator
\begin{equation}
\hat{a}=\sum_{i=1}^N\sqrt{i}\;|i-1\rangle\langle i|\;.
\end{equation}
The quantum-limited attenuator maps $\mathcal{H}_N$ into itself \cite{de2015passive}, and its restriction to $\mathcal{H}_N$ is the channel $e^{t\mathcal{L}}$ generated by the Lindbladian \eqref{Latt}.
In Ref. \cite{de2015passive} it is proven that this quantum channel preserves the set of passive states, and they are its optimal inputs in the sense of Eq. \eqref{optimaldef}.
Here we will show that this last property does no more hold for the restriction to $\mathcal{H}_N\otimes\mathcal{H}_N$ of the two-mode attenuator
\begin{equation}
\mathcal{E}_t:=e^{t\mathcal{L}}\otimes e^{t\mathcal{L}}\;.
\end{equation}
The Hamiltonian is
\begin{equation}
\hat{H}_2=\hat{H}\otimes\hat{\mathbb{I}}+\hat{\mathbb{I}}\otimes\hat{H}=\sum_{i,j=0}^N\left(i+j\right)|i,j\rangle\langle i,j|\;.
\end{equation}
In general there is more than one couple of indices $(i,j)$ with a given sum. Then, $\hat{H}_2$ is degenerate.
However, the two Lindblad operators $\hat{a}\otimes\hat{\mathbb{I}}$ and $\hat{\mathbb{I}}\otimes\hat{a}$ can still induce only jumps between a given energy level and the immediately lower one, and there are no ambiguities in the definition of the passive rearrangement of quantum states with the same degeneracies of the Hamiltonian.
Consider for example
\begin{equation}
\hat{\rho}=\frac{1}{6}\sum_{i+j\leq2} |i,j\rangle\langle i,j|\;,\qquad\mathrm{Tr}\left[\hat{H}_2\;\hat{\rho}\right]=\frac{4}{3}\;.
\end{equation}
It is easy to show that it minimizes the average energy among the states with the same spectrum, i.e. it is passive.
Moreover, there are no other states with the same spectrum and the same average energy, i.e. its passive rearrangement is unique.
Consider instead
\begin{equation}
\hat{\sigma}=\frac{1}{6}\sum_{i=0}^5|0,i\rangle\langle 0,i|\;,\qquad\mathrm{Tr}\left[\hat{H}_2\;\hat{\sigma}\right]=\frac{5}{2}\;,
\end{equation}
that has the same spectrum of $\hat{\rho}$, but it has a higher average energy and it is not passive.
The three largest eigenvalues of $\mathcal{E}_t\left(\hat{\rho}\right)$ are associated with the eigenvectors $|0,0\rangle$, $|0,1\rangle$ and $|1,0\rangle$, and their sum is
\begin{equation}
s_3(t)=1-\frac{e^{-2t}}{2}\;.
\end{equation}
On the other side, the three largest eigenvalues of $\mathcal{E}_t\left(\hat{\sigma}\right)$ are associated with the eigenvectors $|0,0\rangle$, $|0,1\rangle$ and $|0,2\rangle$, and their sum is
\begin{equation}
\tilde{s}_3(t)=1-e^{-3t}\frac{5-6e^{-t}+2e^{-2t}}{2}\;.
\end{equation}
It is then easy to see that for
\begin{equation}
e^{-t}<1-\frac{1}{\sqrt{2}}\;,
\end{equation}
i.e.
\begin{equation}
t>\ln\left(2+\sqrt{2}\right):=t_0\;,
\end{equation}
we have
\begin{equation}
s_3(t)<\tilde{s}_3(t)\;,
\end{equation}
i.e. the passive state $\hat{\rho}$ is not the optimal input.
Let $p_1(t)$ and $\tilde{p}_1(t)$ be the largest eigenvalues of $\mathcal{E}_t\left(\hat{\rho}\right)$ and $\mathcal{E}_t\left(\hat{\sigma}\right)$, respectively.
They are both associated to the eigenvector $|0,0\rangle$, and
\begin{align}
&p_1(t) = \frac{6-8e^{-t}+3e^{-2t}}{6}\\
&\tilde{p}_1(t) = \frac{\left(2-e^{-t}\right)\left(3-3e^{-t}+e^{-2t}\right)\left(1-e^{-t}+e^{-2t}\right)}{6}\;.
\end{align}
For any $t>0$
\begin{equation}
p_1(t)>\tilde{p}_1(t)\;,
\end{equation}
so that $\hat{\sigma}$ is not the optimal input, and for $t>t_0$ no majorization relation holds between $\mathcal{E}_t\left(\hat{\rho}\right)$ and $\mathcal{E}_t\left(\hat{\sigma}\right)$.

\subsection{Two-qubit lossy channel} \label{sec2step}
We consider a quantum lossy channel acting on the quantum system of two qubits with two possible choices for the Hamiltonian, and we show that passive states are not the optimal inputs in the sense of \eqref{optimaldef}.
In one case (Section \ref{+jumps}) the Hamiltonian is non-degenerate, but the channel involves quantum jumps of more than one energy step.
In the other case (Section \ref{degH}), only quantum jumps of one energy step are allowed, but the Hamiltonian becomes degenerate.

Let us consider the Hilbert space of two distinguishable spins with Hamiltonian
\begin{equation}\label{Hspin}
\hat{H}=E_1\;|1\rangle\langle1|\otimes\hat{\mathbb{I}}+E_2\;\hat{\mathbb{I}}\otimes|1\rangle\langle 1|\;.
\end{equation}
We notice that $\hat{H}$ is not symmetric under the exchange of the two spins, i.e. the spins are different, though the same Hilbert space $\mathbb{C}^2$ is associated to both of them.
Let us suppose that
\begin{equation}
0< E_2\leq E_1\;,
\end{equation}
so that the eigenvectors of $\hat{H}$ are, in order of increasing energy,
\begin{eqnarray}
\hat{H}|0,0\rangle &=& 0\nonumber\\
\hat{H}|0,1\rangle &=& E_2|0,1\rangle\nonumber\\
\hat{H}|1,0\rangle &=& E_1|1,0\rangle\nonumber\\
\hat{H}|1,1\rangle &=& (E_1+E_2)|1,1\rangle\;,
\end{eqnarray}
with the only possible degeneracy between $|0,1\rangle$ and $|1,0\rangle$ if $E_1=E_2$.

Let $\mathcal{L}$ be the generator of the form \eqref{Ldef} with the two Lindblad operators
\begin{eqnarray}
\hat{L}_1&=& |0,0\rangle\langle1,0|\nonumber\\
\hat{L}_2 &=& |0,0\rangle\langle0,1|+\sqrt{2}\;|0,1\rangle\langle 1,1|\;,
\end{eqnarray}
and let
\begin{equation}
\mathcal{E}_t=e^{t\mathcal{L}}\;,\qquad t\geq0\;,
\end{equation}
be the associated quantum channel.

\subsubsection{Jumps of more than one energy step}\label{+jumps}
If $E_2<E_1$ the Hamiltonian \eqref{Hspin} is non-degenerate, but the Lindblad operator $\hat{L}_2$ can induce a transition from $|1,1\rangle$ to $|0,1\rangle$, that are not consecutive eigenstates.

For simplicity, we parameterize a state diagonal in the Hamiltonian eigenbasis with
\begin{equation}\label{rhopar}
\hat{\rho}=\sum_{i,j=0}^1 p_{ij}\;|i,j\rangle\langle i,j|\;.
\end{equation}
First, let
\begin{equation}\label{rho0}
\hat{\rho}^{(0)}(t)=\mathcal{E}_t\left(\frac{\hat{\mathbb{I}}}{4}\right)
\end{equation}
be the output of the channel applied to the maximally mixed state.
Then, we can compute
\begin{eqnarray}\label{pij}
p^{(0)}_{00}(t) &=& 1-e^{-t}+\frac{e^{-2t}}{4}\nonumber\\
p^{(0)}_{01}(t) &=& e^{-t}\;\frac{3-2e^{-t}}{4}\nonumber\\
p^{(0)}_{10}(t) &=& \frac{e^{-t}}{4}\nonumber\\
p^{(0)}_{11}(t) &=& \frac{e^{-2t}}{4}\;.
\end{eqnarray}
It is easy to check that, for any $t>0$,
\begin{equation}\label{pij>}
p^{(0)}_{00}(t)>p^{(0)}_{01}(t)>p^{(0)}_{10}(t)>p^{(0)}_{11}(t)\;,
\end{equation}
so that $\hat{\rho}^{(0)}(t)$ is passive, and the channel $\mathcal{E}_t$ satisfies the hypothesis of Lemma \ref{passl}.
Let us instead compare
\begin{align}\label{rho1}
&\hat{\rho}^{(1)}(t) = \mathcal{E}_t\left(\frac{|0,0\rangle\langle0,0|+|0,1\rangle\langle0,1|+|1,0\rangle\langle1,0|}{3}\right)\\
&\hat{\rho}^{(2)}(t) = \mathcal{E}_t\left(\frac{|0,0\rangle\langle0,0|+|0,1\rangle\langle0,1|+|1,1\rangle\langle1,1|}{3}\right)\;.\label{rho2}
\end{align}
It is easy to see that $\hat{\rho}^{(1)}(0)$ is passive, while $\hat{\rho}^{(2)}(0)$ is not, and they have the same spectrum.
Moreover, there are no other states with the same spectrum and the same average energy of $\hat{\rho}^{(1)}(0)$, i.e. its passive rearrangement is unique.
We can now compute
\begin{align}\label{p12t}
&p^{(1)}_{00}(t) = 1-\frac{2}{3}e^{-t}\qquad && p^{(2)}_{00}(t) = 1-e^{-t}+\frac{e^{-2t}}{3}\nonumber\\
&p^{(1)}_{01}(t) = \frac{e^{-t}}{3}\qquad && p^{(2)}_{01}(t) = e^{-t}\left(1-\frac{2}{3}e^{-t}\right)\nonumber\\
&p^{(1)}_{10}(t) = \frac{e^{-t}}{3}\qquad && p^{(2)}_{10}(t) =0\nonumber\\
&p^{(1)}_{11}(t) = 0\qquad && p^{(2)}_{11}(t) =\frac{e^{-2t}}{3}\;.
\end{align}
It is easy to see that for any $t>0$
\begin{align}\label{p12>}
&p^{(1)}_{00}(t)>p^{(1)}_{01}(t)=p^{(1)}_{10}(t)\nonumber\\
&p^{(2)}_{00}(t)>p^{(2)}_{01}(t)>p^{(2)}_{11}(t)\;,
\end{align}
so that $\hat{\rho}^{(1)}(t)$ remains always passive.
However, on one hand
\begin{equation}
p^{(1)}_{00}(t)>p^{(2)}_{00}(t)\;,
\end{equation}
but on the other hand
\begin{equation}
p^{(1)}_{00}(t)+p^{(1)}_{01}(t)<p^{(2)}_{00}(t)+p^{(2)}_{01}(t)\;,
\end{equation}
so that no majorization relation can exist between $\hat{\rho}^{(1)}(t)$ and $\hat{\rho}^{(2)}(t)$.

\subsubsection{Degenerate Hamiltonian}\label{degH}
If $E_1=E_2$, the eigenstates $|0,1\rangle$ and $|1,0\rangle$ of the Hamiltonian \eqref{Hspin} become degenerate, but both $\hat{L}_1$ and $\hat{L}_2$ induce only transitions between consecutive energy levels.

We use the parametrization \eqref{rhopar} as before.
Let $\hat{\rho}^{(0)}(t)$ be the output of the channel applied to the maximally mixed state as in \eqref{rho0}.
Since the generator $\mathcal{L}$ is the same of Section \ref{+jumps}, the probabilities $p^{(0)}_{ij}(t)$, $i,j=0,1$, are still given by \eqref{pij}.
Eq. \eqref{pij>} still holds for any $t>0$, so that $\hat{\rho}^{(0)}(t)$ is passive, and the channel $\mathcal{E}_t$ satisfies the hypothesis of Lemma \ref{passl}.

Let us instead compare $\hat{\rho}^{(1)}(t)$ and $\hat{\rho}^{(2)}(t)$ defined as in \eqref{rho1} and \eqref{rho2}, respectively.
The state $\hat{\rho}^{(1)}(0)$ is passive, while $\hat{\rho}^{(2)}(0)$ is not, and they have the same spectrum.
Moreover, there are no other states with the same spectrum and the same average energy of $\hat{\rho}^{(1)}(0)$, i.e. its passive rearrangement is unique.
The probabilities $p^{(1)}_{ij}(t)$ and $p^{(2)}_{ij}(t)$, $i,j=0,1$, are still given by \eqref{p12t}.
Eq. \eqref{p12>} still holds for any $t>0$, and $\hat{\rho}^{(1)}(t)$ remains always passive.
However, on one hand
\begin{equation}
p^{(1)}_{00}(t)>p^{(2)}_{00}(t)\;,
\end{equation}
but on the other hand
\begin{equation}
p^{(1)}_{00}(t)+p^{(1)}_{01}(t)<p^{(2)}_{00}(t)+p^{(2)}_{01}(t)\;,
\end{equation}
so that no majorization relation can exist between $\hat{\rho}^{(1)}(t)$ and $\hat{\rho}^{(2)}(t)$.

\subsection{Optimal states for a finite-temperature two-level system are nonclassical}\label{secfin}
In this section we show that at finite temperature, already for a two-level system the optimal states are no more passive, and include coherent superpositions of the energy eigenstates.

An intuitive explanation is that a dissipator with only energy-raising Lindblad operators keeps fixed the maximum-energy eigenstate, that is hence optimal for the generated channel.
Then, it is natural to expect that the optimal pure state in the presence of both energy-lowering and energy-raising Lindblad operators will interpolate between the ground and the maximum energy state, and will hence be a coherent superposition of different eigenstates of the Hamiltonian.

The simplest example is a two-level system with Hamiltonian
\begin{equation}
\hat{H}=\frac{1}{2}E_0\;\hat{\sigma}_z=\frac{E_0}{2}\;|1\rangle\langle 1|-\frac{E_0}{2}\;|0\rangle\langle0|\;,\qquad E_0>0\;,
\end{equation}
undergoing the quantum optical master equation \cite{breuer2007theory}, describing the weak coupling with a thermal bath of one mode of bosonic excitations in the rotating-wave approximation.
This is the simplest extension of the evolutions considered in the previous section to an interaction with a finite-temperature bath.

Its generator is
\begin{eqnarray}\label{qubitL}
\mathcal{L}\left(\hat{\rho}\right) &=& \gamma_0(N+1)\left(\hat{\sigma}_-\;\hat{\rho}\;\hat{\sigma}_+-\frac{1}{2}\left\{\hat{\sigma}_+\hat{\sigma}_-,\;\hat{\rho}\right\}\right)+\nonumber\\
&&+\gamma_0N\left(\hat{\sigma}_+\;\hat{\rho}\;\hat{\sigma}_--\frac{1}{2}\left\{\hat{\sigma}_-\hat{\sigma}_+,\;\hat{\rho}\right\}\right)\;,
\end{eqnarray}
where
\begin{equation}
\hat{\sigma}_\pm=\frac{\hat{\sigma}_x\pm i\hat{\sigma}_y}{2}
\end{equation}
are the ladder operators, $\gamma_0>0$ is the coupling strength and $N>0$ is the average number of photons or phonons in the bosonic mode of the bath coupled to the system. Notice also that
for $N=0$ the process becomes a lossy map fulfilling the condition discussed at the beginning of Sec.~\ref{secopt}.

We will now show that, for the quantum channel associated to the master equation \eqref{qubitL}, the output generated by a certain coherent superposition of the two energy eigenstates majorizes the output generated by any other state.

It is convenient to use the Bloch representation
\begin{equation}\label{bloch}
\hat{\rho}=\frac{\hat{\mathbb{I}}+x\;\hat{\sigma}_x+y\;\hat{\sigma}_y+z\;\hat{\sigma}_z}{2}\;,\qquad x^2+y^2+z^2\leq1\;.
\end{equation}
The master equation \eqref{qubitL} induces the differential equations
\begin{equation}\label{vdot}
\frac{dx}{dt} = -\frac{\gamma}{2}\;x\;,\qquad \frac{dy}{dt} = -\frac{\gamma}{2}\;y\;,\qquad \frac{dz}{dt} = -\gamma\left(z-z_\infty\right)\;,
\end{equation}
where
\begin{equation}
\gamma=\gamma_0(2N+1)\qquad\text{and}\qquad z_\infty=-\frac{1}{2N+1}\;.
\end{equation}
The solution of \eqref{vdot} is
\begin{eqnarray}
x(t) &=& e^{-\frac{\gamma}{2}t}\;x_0\;,\nonumber\\
y(t) &=& e^{-\frac{\gamma}{2}t}\;y_0\;,\nonumber\\
z(t) &=& z_\infty+e^{-\gamma t}\left(z_0-z_\infty\right)\;,
\end{eqnarray}
and its asymptotic state is the canonical state with inverse temperature $\beta$
\begin{equation}
\hat{\rho}_\infty=\frac{e^\frac{\beta E_0}{2}\;|0\rangle\langle0|+e^{-\frac{\beta E_0}{2}}\;|1\rangle\langle 1|}{2\cosh\frac{\beta E_0}{2}}\;,
\end{equation}
satisfying
\begin{equation}
z_\infty=-\tanh\frac{\beta\;E_0}{2}\;.
\end{equation}
Since the density matrix of a two-level system has only two eigenvalues, the purity is a sufficient criterion for majorization, i.e. for any two quantum states $\hat{\rho}$ and $\hat{\sigma}$,
\begin{equation}
\hat{\rho}\prec\hat{\sigma}\qquad\text{iff}\qquad\mathrm{Tr}\;\hat{\rho}^2\leq\mathrm{Tr}\;\hat{\sigma}^2\;.
\end{equation}
We recall that in the Bloch representation \eqref{bloch}
\begin{equation}
\mathrm{Tr}\;\hat{\rho}^2=\frac{1+x^2+y^2+z^2}{2}\;.
\end{equation}
We have then
\begin{align}\label{purity}
\mathrm{Tr}\;{\hat{\rho}(t)}^2 &=& \frac{1+e^{-\gamma t}\left(x_0^2+y_0^2+z_0^2\right)}{2}+\qquad\qquad\qquad\qquad\nonumber\\
&&+\frac{1-e^{-\gamma t}}{2}\left(z_\infty^2-e^{-\gamma t}\left(z_0-z_\infty\right)^2\right)\;.\qquad\quad
\end{align}
The right-hand side of \eqref{purity} is maximized by
\begin{equation}
x_0^2+y_0^2 = 1-z_\infty^2\qquad\text{and}\qquad z_0 = z_\infty\;,
\end{equation}
i.e. when the initial state is a pure coherent superposition of the energy eigenstates $|0\rangle$ and $|1\rangle$ with the same average energy of the asymptotic state:
\begin{equation}
|\psi\rangle=e^{i\varphi_0}\sqrt{\frac{1-z_\infty}{2}}\;|0\rangle+e^{i\varphi_1}\sqrt{\frac{1+z_\infty}{2}}\;|1\rangle\;,
\end{equation}
where $\varphi_0$ and $\varphi_1$ are arbitrary real phases.

\section{Conclusions} \label{seccon}
In this paper we have extended the proof of the optimality of passive states of Ref. \cite{de2015passive} to a large class of lossy channels, showing that they preserve the set of passive states, that are the optimal inputs in the sense that the output generated by a passive state majorizes the output generated by any other state with the same spectrum.
Then, thanks to the equivalent definition of majorization in terms of random unitary operations \eqref{majru}, the output generated by a passive state minimizes any concave functional among the outputs generated by any unitary equivalent state.
Since the class of concave functionals includes the von Neumann and all the R\'enyi entropies, the solution to any entropic optimization problem has to be found among passive states.
This result can then lead to entropic inequalities on the output of a lossy channel, and can be crucial in the determination of its information capacity.
Moreover, in the context of quantum thermodynamics this result can be useful to determine which quantum states can be obtained from an initial state with a given spectrum in a resource theory with lossy channels among the allowed operations.

The optimality of passive states crucially depends on the assumptions of non-degeneracy of the Hamiltonian, quantum jumps of only one energy step and zero temperature.
Indeed, the two-mode bosonic Gaussian quantum-limited attenuator provides a counterexample with degenerate Hamiltonian.
Moreover, two-qubit systems can provide counterexamples both with degenerate Hamiltonian or with quantum jumps of more than one energy step.
Finally, at finite temperature this optimality property fails already for a two-level system, where the best input is a coherent superposition of the two energy eigenstates.
This shows that even the quantum channels that naturally arise from a weak interaction with a thermal bath can have a very complex entropic behaviour, and that coherence can play a crucial role in the optimal encoding of information.

\appendix
\section{Auxiliary lemmata}
\subsection{Passivity of the evolved maximally mixed state}\label{passt}
\begin{lem}\label{passl}
Let $\mathcal{L}$ be a Lindblad generator such that for any $t\geq0$ the operator $e^{t\mathcal{L}}\left(\hat{\mathbb{I}}\right)$ is passive.
Then, also $\mathcal{L}\left(\hat{\mathbb{I}}\right)$ is passive.
\begin{proof}
Recalling the Hamiltonian eigenbasis \eqref{Hpass}, for any $t\geq0$ it must hold
\begin{equation}
e^{t\mathcal{L}}\left(\hat{\mathbb{I}}\right)=\sum_{i=1}^d c_i(t)\;|i\rangle\langle i|
\end{equation}
with
\begin{equation}
c_1(t)\geq\ldots\geq c_d(t)\;,\qquad c_1(0)=\ldots c_d(0)=1\;,
\end{equation}
and each $c_i(t)$ is an analytic function of $t$.
It follows that
\begin{equation}
c_1'(0)\geq\ldots\geq c_d'(0)\;.
\end{equation}
However, we also have
\begin{equation}
\mathcal{L}\left(\hat{\mathbb{I}}\right)=\left.\frac{d}{dt}e^{t\mathcal{L}}\left(\hat{\mathbb{I}}\right)\right|_{t=0}=\sum_{i=1}^d c_i'(0)\;|i\rangle\langle i|\;,
\end{equation}
hence the thesis.
\end{proof}
\end{lem}

\subsection{Proof of Lemma \ref{deg}}\label{proofdeg}
The matrix elements of the operator $e^{t\mathcal{L}}\left(\hat{\rho}\right)$ are analytic functions of $t$.
The spectrum of $\hat{\rho}(t)$ is degenerate iff the function
\begin{equation}
\phi(t)=\prod_{i\neq j}\left(p_i(t)-p_j(t)\right)
\end{equation}
vanishes.
This function is a symmetric polynomial in the eigenvalues of $\hat{\rho}(t)=e^{t\mathcal{L}}\left(\hat{\rho}\right)$.
Then, for the Fundamental Theorem of Symmetric Polynomials (see e.g Theorem 3 in Chapter 7 of \cite{cox2015ideals}), $\phi(t)$ can be written as a polynomial in the elementary symmetric polynomials in the eigenvalues of $\hat{\rho}(t)$.
However, these polynomials coincide with the coefficients of the characteristic polynomial of $\hat{\rho}(t)$, that are in turn polynomials in its matrix elements.
It follows that $\phi(t)$ can be written as a polynomial in the matrix elements of the operator $\hat{\rho}(t)$.
Since each of these matrix element is an analytic function of $t$, also $\phi(t)$ is analytic.
Since by hypothesis the spectrum of $\hat{\rho}(0)$ is non-degenerate, $\phi$ cannot be identically zero, and its zeroes are isolated points.

\subsection{Proof of Lemma \ref{lemma1}}\label{prooflemma1}
The matrix elements of the operator $e^{t\mathcal{L}}\left(\hat{\rho}\right)$ are analytic (and hence continuous and differentiable) functions of $t$.
Then for Weyl's Perturbation Theorem $p(t)$ is continuous in $t$, and also $s(t)$ is continuous (see e.g. Corollary III.2.6 and the discussion at the beginning of Chapter VI  of \cite{bhatia2013matrix}).
Let $\hat{\rho}(t_0)$ have non-degenerate spectrum.
Then, $\hat{\rho}(t)$ has non-degenerate spectrum for any $t$ in a suitable neighbourhood of $t_0$.
In this neighbourhood, we can diagonalize $\hat{\rho}(t)$ with
\begin{equation}
\hat{\rho}(t)=\sum_{i=1}^d p_i(t) |\psi_i(t)\rangle\langle\psi_i(t)|\;,
\end{equation}
where the eigenvalues in decreasing order $p_i(t)$ are differentiable functions of $t$ (see Theorem 6.3.12 of \cite{horn2012matrix}).
We then have
\begin{equation}
\frac{d}{dt}p_i(t)=\langle\psi_i(t)|\mathcal{L}\left(\hat{\rho}(t)\right)|\psi_i(t)\rangle\;,\qquad i=1,\ldots,d\;,
\end{equation}
and
\begin{equation}
\frac{d}{dt}s_n(t)=\mathrm{Tr}\left[\hat{\Pi}_n(t)\;\mathcal{L}\left(\hat{\rho}(t)\right)\right]\;,
\end{equation}
where
\begin{equation}
\hat{\Pi}_n(t)=\sum_{i=1}^n|\psi_i(t)\rangle\langle\psi_i(t)|\;.
\end{equation}
We can write
\begin{equation}
\hat{\rho}(t)=\sum_{n=1}^d d_n(t)\;\hat{\Pi}_n(t)\;,
\end{equation}
where
\begin{equation}
d_n(t)=p_n(t)-p_{n+1}(t)\geq0\;,
\end{equation}
and for simplicity we have set $p_{d+1}(t)=0$, so that
\begin{equation}\label{sndot}
\frac{d}{dt}s_n(t)=\sum_{k=1}^d d_k(t)\;\mathrm{Tr}\left[\hat{\Pi}_n(t)\;\mathcal{L}\left(\hat{\Pi}_k(t)\right)\right]\;.
\end{equation}
We have now
\begin{align}\label{PLext}
&\mathrm{Tr}\left[\hat{\Pi}_n(t)\;\mathcal{L}\left(\hat{\Pi}_k(t)\right)\right]=\nonumber\\
&=\sum_\alpha\mathrm{Tr}\left[\hat{\Pi}_n(t)\;\hat{L}_\alpha\;\hat{\Pi}_k(t)\;\hat{L}_\alpha^\dag-\hat{\Pi}_{k\land n}(t)\;\hat{L}_\alpha^\dag\hat{L}_\alpha\right]\;,
\end{align}
where $k\land n=\min(k,n)$ and we have used that
\begin{equation}
\hat{\Pi}_n(t)\;\hat{\Pi}_k(t)=\hat{\Pi}_k(t)\;\hat{\Pi}_n(t)=\hat{\Pi}_{k\land n}(t)\;.
\end{equation}
\begin{itemize}
  \item Let us suppose $n\geq k$.
  Using that $\hat{\Pi}_n(t)\leq\hat{\mathbb{I}}$ in the first term of \eqref{PLext}, we get
  \begin{equation}\label{n>k}
  \mathrm{Tr}\left[\hat{\Pi}_n(t)\;\mathcal{L}\left(\hat{\Pi}_k(t)\right)\right]\leq0\;.
  \end{equation}
  On the other hand, recalling the structure of the Lindblad operators \eqref{dephase} and \eqref{jump}, for any $\alpha$ the support of $\hat{L}_\alpha\,\hat{\Pi}_k^\downarrow\,\hat{L}_\alpha^\dag$ is contained into the support of $\hat{\Pi}_{k}^\downarrow$, and hence into the one of $\hat{\Pi}_n^\downarrow$, and we have also
  \begin{equation}\label{n>k*}
  \mathrm{Tr}\left[\hat{\Pi}^\downarrow_n\;\mathcal{L}\left(\hat{\Pi}_k^\downarrow\right)\right]=0\;.
  \end{equation}
  \item Let us now suppose $k>n$.
  Using that $\hat{\Pi}_k(t)\leq\hat{\mathbb{I}}$ in the first term of \eqref{PLext}, we get
  \begin{align}\label{k>n}
  \mathrm{Tr}\left[\hat{\Pi}_n(t)\;\mathcal{L}\left(\hat{\Pi}_k(t)\right)\right] &\leq \mathrm{Tr}\left[\hat{\Pi}_n(t)\;\mathcal{L}\left(\hat{\mathbb{I}}\right)\right] \leq\nonumber\\
  &\leq \mathrm{Tr}\left[\hat{\Pi}_n^\downarrow\;\mathcal{L}\left(\hat{\mathbb{I}}\right)\right]=\lambda_n\;,
  \end{align}
  where in the last step we have used Ky Fan's maximum principle (Lemma \ref{sumeig}) and the passivity of $\mathcal{L}\left(\hat{\mathbb{I}}\right)$.
  On the other hand, from \eqref{dephase} and \eqref{jump} the support of $\hat{L}_\alpha^\dag\,\hat{\Pi}_n^\downarrow\,\hat{L}_\alpha$ is contained into the support of $\hat{\Pi}_{n+1}^\downarrow$, and hence into the one of $\hat{\Pi}_k^\downarrow$, and we have also
  \begin{equation}\label{n<k*}
  \mathrm{Tr}\left[\hat{\Pi}_n^\downarrow\;\mathcal{L}\left(\hat{\Pi}_k^\downarrow\right)\right]=\lambda_n\;.
  \end{equation}
\end{itemize}
Plugging \eqref{n>k} and \eqref{k>n} into \eqref{sndot}, we get
\begin{equation}
\frac{d}{dt}s_n(t)\leq\lambda_n\;p_{n+1}(t)=\lambda_n\left(s_{n+1}(t)-s_n(t)\right)\;.
\end{equation}
From \eqref{n>k*} and \eqref{n<k*} we get instead
\begin{equation}
\frac{d}{dt}s_n^\downarrow(t)=\lambda_n\;p_{n+1}^\downarrow(t)=\lambda_n\left(s^\downarrow_{n+1}(t)-s_n^\downarrow(t)\right)\;.
\end{equation}
See Lemma \ref{lambdan} for the positivity of the coefficients $\lambda_n$.

\subsection{Proof of Lemma \ref{lemma2}}\label{prooflemma2}
Since the quantum channel $e^{t\mathcal{L}}$ is trace-preserving, we have
\begin{equation}
s_d(t)=\mathrm{Tr}\;\hat{\rho}(t)=1=s_d^\downarrow(t)\;.
\end{equation}
We will use induction on $n$ in the reverse order: suppose to have proved
\begin{equation}
s_{n+1}(t)\leq s_{n+1}^\downarrow(t)\;.
\end{equation}
Since $\lambda_n\geq0$ for Lemma \ref{lambdan}, we have from \eqref{sdot}
\begin{equation}
\frac{d}{dt}s_n(t)\leq\lambda_n\left(s_{n+1}^\downarrow(t)-s_n(t)\right)\;,
\end{equation}
while
\begin{equation}
\frac{d}{dt}s_n^\downarrow(t)=\lambda_n\left(s_{n+1}^\downarrow(t)-s_n^\downarrow(t)\right)\;.
\end{equation}
Defining
\begin{equation}
f_n(t)=s_n^\downarrow(t)-s_n(t)\;,
\end{equation}
we have $f_n(0)=0$, and
\begin{equation}
\frac{d}{dt}f_n(t)\geq-\lambda_n\;f_n(t)\;.
\end{equation}
This can be rewritten as
\begin{equation}
e^{-\lambda_nt}\;\frac{d}{dt}\left(e^{\lambda_nt}\;f_n(t)\right)\geq0\;,
\end{equation}
and implies
\begin{equation}
f_n(t)\geq0\;.
\end{equation}

\subsection{Ky Fan's Maximum Principle}
\begin{lem}[Ky Fan's Maximum Principle]\label{sumeig}
Let $\hat{X}$ be a self-adjoint operator
with eigenvalues $x_1\geq\ldots\geq x_d$, and let $\hat{P}$ be a projector of rank $n$.
Then
\begin{equation}\label{TrPiX}
\mathrm{Tr}\left[\hat{P}\;\hat{X}\right]\leq\sum_{i=1}^n x_i\;.
\end{equation}
\begin{proof}
See \cite{bhatia2013matrix,fan1951maximum} or \cite{de2015passive}.
\end{proof}
\end{lem}

\subsection{Proof of Lemma \ref{lambdan}}\label{prooflambdan}
\begin{lem}\label{lambdan}
$\lambda_n\geq0$ for $n=1,\,\ldots,\,d$.
\begin{proof}
For Ky Fan's maximum principle (Lemma \ref{sumeig}), for any unitary $\hat{U}$
\begin{equation}\label{lambdaU}
\lambda_n = \mathrm{Tr}\left[\hat{\Pi}_n^\downarrow\;\mathcal{L}\left(\hat{\mathbb{I}}\right)\right]\geq \mathrm{Tr}\left[\hat{U}\;\hat{\Pi}_n^\downarrow\;\hat{U}^\dag\;\mathcal{L}\left(\hat{\mathbb{I}}\right)\right]\;.
\end{equation}
The thesis easily follows taking the average over the Haar measure $\mu$ of the right-hand side of \eqref{lambdaU}, since
\begin{equation}
\int\hat{U}^\dag\;\mathcal{L}\left(\hat{\mathbb{I}}\right)\;\hat{U}\;d\mu\left(\hat{U}\right)= \frac{\hat{\mathbb{I}}}{d}\;\mathrm{Tr}\left[\mathcal{L}\left(\hat{\mathbb{I}}\right)\right]=0\;.
\end{equation}
\end{proof}
\end{lem}

\bibliography{biblio}

\begin{thebibliography}{31}%
\makeatletter
\providecommand \@ifxundefined [1]{%
 \@ifx{#1\undefined}
}%
\providecommand \@ifnum [1]{%
 \ifnum #1\expandafter \@firstoftwo
 \else \expandafter \@secondoftwo
 \fi
}%
\providecommand \@ifx [1]{%
 \ifx #1\expandafter \@firstoftwo
 \else \expandafter \@secondoftwo
 \fi
}%
\providecommand \natexlab [1]{#1}%
\providecommand \enquote  [1]{``#1''}%
\providecommand \bibnamefont  [1]{#1}%
\providecommand \bibfnamefont [1]{#1}%
\providecommand \citenamefont [1]{#1}%
\providecommand \href@noop [0]{\@secondoftwo}%
\providecommand \href [0]{\begingroup \@sanitize@url \@href}%
\providecommand \@href[1]{\@@startlink{#1}\@@href}%
\providecommand \@@href[1]{\endgroup#1\@@endlink}%
\providecommand \@sanitize@url [0]{\catcode `\\12\catcode `\$12\catcode
  `\&12\catcode `\#12\catcode `\^12\catcode `\_12\catcode `\%12\relax}%
\providecommand \@@startlink[1]{}%
\providecommand \@@endlink[0]{}%
\providecommand \url  [0]{\begingroup\@sanitize@url \@url }%
\providecommand \@url [1]{\endgroup\@href {#1}{\urlprefix }}%
\providecommand \urlprefix  [0]{URL }%
\providecommand \Eprint [0]{\href }%
\providecommand \doibase [0]{http://dx.doi.org/}%
\providecommand \selectlanguage [0]{\@gobble}%
\providecommand \bibinfo  [0]{\@secondoftwo}%
\providecommand \bibfield  [0]{\@secondoftwo}%
\providecommand \translation [1]{[#1]}%
\providecommand \BibitemOpen [0]{}%
\providecommand \bibitemStop [0]{}%
\providecommand \bibitemNoStop [0]{.\EOS\space}%
\providecommand \EOS [0]{\spacefactor3000\relax}%
\providecommand \BibitemShut  [1]{\csname bibitem#1\endcsname}%
\let\auto@bib@innerbib\@empty
\bibitem [{\citenamefont {Pusz}\ and\ \citenamefont
  {Woronowicz}(1978)}]{pusz1978passive}%
  \BibitemOpen
  \bibfield  {author} {\bibinfo {author} {\bibfnamefont {W.}~\bibnamefont
  {Pusz}}\ and\ \bibinfo {author} {\bibfnamefont {S.}~\bibnamefont
  {Woronowicz}},\ }\href@noop {} {\bibfield  {journal} {\bibinfo  {journal}
  {Communications in Mathematical Physics}\ }\textbf {\bibinfo {volume} {58}},\
  \bibinfo {pages} {273} (\bibinfo {year} {1978})}\BibitemShut {NoStop}%
\bibitem [{\citenamefont {Lenard}(1978)}]{lenard1978thermodynamical}%
  \BibitemOpen
  \bibfield  {author} {\bibinfo {author} {\bibfnamefont {A.}~\bibnamefont
  {Lenard}},\ }\href@noop {} {\bibfield  {journal} {\bibinfo  {journal}
  {Journal of Statistical Physics}\ }\textbf {\bibinfo {volume} {19}},\
  \bibinfo {pages} {575} (\bibinfo {year} {1978})}\BibitemShut {NoStop}%
\bibitem [{\citenamefont {Janzing}(2006)}]{janzing2006computational}%
  \BibitemOpen
  \bibfield  {author} {\bibinfo {author} {\bibfnamefont {D.}~\bibnamefont
  {Janzing}},\ }\href@noop {} {\bibfield  {journal} {\bibinfo  {journal}
  {Journal of statistical physics}\ }\textbf {\bibinfo {volume} {122}},\
  \bibinfo {pages} {531} (\bibinfo {year} {2006})}\BibitemShut {NoStop}%
\bibitem [{\citenamefont {Vinjanampathy}\ and\ \citenamefont
  {Anders}(2015)}]{vinjanampathy2015quantum}%
  \BibitemOpen
  \bibfield  {author} {\bibinfo {author} {\bibfnamefont {S.}~\bibnamefont
  {Vinjanampathy}}\ and\ \bibinfo {author} {\bibfnamefont {J.}~\bibnamefont
  {Anders}},\ }\href@noop {} {\bibfield  {journal} {\bibinfo  {journal} {arXiv
  preprint arXiv:1508.06099}\ } (\bibinfo {year} {2015})}\BibitemShut {NoStop}%
\bibitem [{\citenamefont {Goold}\ \emph {et~al.}(2016)\citenamefont {Goold},
  \citenamefont {Huber}, \citenamefont {Riera}, \citenamefont {del Rio},\ and\
  \citenamefont {Skrzypczyk}}]{goold2015role}%
  \BibitemOpen
  \bibfield  {author} {\bibinfo {author} {\bibfnamefont {J.}~\bibnamefont
  {Goold}}, \bibinfo {author} {\bibfnamefont {M.}~\bibnamefont {Huber}},
  \bibinfo {author} {\bibfnamefont {A.}~\bibnamefont {Riera}}, \bibinfo
  {author} {\bibfnamefont {L.}~\bibnamefont {del Rio}}, \ and\ \bibinfo
  {author} {\bibfnamefont {P.}~\bibnamefont {Skrzypczyk}},\ }\href@noop {}
  {\bibfield  {journal} {\bibinfo  {journal} {Journal of Physics A:
  Mathematical and Theoretical}\ }\textbf {\bibinfo {volume} {49}},\ \bibinfo
  {pages} {143001} (\bibinfo {year} {2016})}\BibitemShut {NoStop}%
\bibitem [{\citenamefont {Marshall}\ \emph {et~al.}(2010)\citenamefont
  {Marshall}, \citenamefont {Olkin},\ and\ \citenamefont
  {Arnold}}]{marshall2010inequalities}%
  \BibitemOpen
  \bibfield  {author} {\bibinfo {author} {\bibfnamefont {A.}~\bibnamefont
  {Marshall}}, \bibinfo {author} {\bibfnamefont {I.}~\bibnamefont {Olkin}}, \
  and\ \bibinfo {author} {\bibfnamefont {B.}~\bibnamefont {Arnold}},\
  }\href@noop {} {\emph {\bibinfo {title} {Inequalities: Theory of Majorization
  and Its Applications}}},\ Springer Series in Statistics\ (\bibinfo
  {publisher} {Springer New York},\ \bibinfo {year} {2010})\BibitemShut
  {NoStop}%
\bibitem [{\citenamefont {Gour}\ \emph {et~al.}(2015)\citenamefont {Gour},
  \citenamefont {M{\"u}ller}, \citenamefont {Narasimhachar}, \citenamefont
  {Spekkens},\ and\ \citenamefont {Halpern}}]{gour2015resource}%
  \BibitemOpen
  \bibfield  {author} {\bibinfo {author} {\bibfnamefont {G.}~\bibnamefont
  {Gour}}, \bibinfo {author} {\bibfnamefont {M.~P.}\ \bibnamefont
  {M{\"u}ller}}, \bibinfo {author} {\bibfnamefont {V.}~\bibnamefont
  {Narasimhachar}}, \bibinfo {author} {\bibfnamefont {R.~W.}\ \bibnamefont
  {Spekkens}}, \ and\ \bibinfo {author} {\bibfnamefont {N.~Y.}\ \bibnamefont
  {Halpern}},\ }\href@noop {} {\bibfield  {journal} {\bibinfo  {journal}
  {Physics Reports}\ } (\bibinfo {year} {2015})}\BibitemShut {NoStop}%
\bibitem [{\citenamefont {Horodecki}\ and\ \citenamefont
  {Oppenheim}(2013)}]{horodecki2013fundamental}%
  \BibitemOpen
  \bibfield  {author} {\bibinfo {author} {\bibfnamefont {M.}~\bibnamefont
  {Horodecki}}\ and\ \bibinfo {author} {\bibfnamefont {J.}~\bibnamefont
  {Oppenheim}},\ }\href@noop {} {\bibfield  {journal} {\bibinfo  {journal}
  {Nature communications}\ }\textbf {\bibinfo {volume} {4}} (\bibinfo {year}
  {2013})}\BibitemShut {NoStop}%
\bibitem [{\citenamefont {Nielsen}(1999)}]{nielsen1999conditions}%
  \BibitemOpen
  \bibfield  {author} {\bibinfo {author} {\bibfnamefont {M.~A.}\ \bibnamefont
  {Nielsen}},\ }\href@noop {} {\bibfield  {journal} {\bibinfo  {journal}
  {Physical Review Letters}\ }\textbf {\bibinfo {volume} {83}},\ \bibinfo
  {pages} {436} (\bibinfo {year} {1999})}\BibitemShut {NoStop}%
\bibitem [{\citenamefont {Nielsen}\ and\ \citenamefont
  {Vidal}(2001)}]{nielsen2001majorization}%
  \BibitemOpen
  \bibfield  {author} {\bibinfo {author} {\bibfnamefont {M.~A.}\ \bibnamefont
  {Nielsen}}\ and\ \bibinfo {author} {\bibfnamefont {G.}~\bibnamefont
  {Vidal}},\ }\href@noop {} {\bibfield  {journal} {\bibinfo  {journal} {Quantum
  Information \& Computation}\ }\textbf {\bibinfo {volume} {1}},\ \bibinfo
  {pages} {76} (\bibinfo {year} {2001})}\BibitemShut {NoStop}%
\bibitem [{\citenamefont {Giovannetti}\ \emph {et~al.}(2014)\citenamefont
  {Giovannetti}, \citenamefont {Garc{\'\i}a-Patr{\'o}n}, \citenamefont {Cerf},\
  and\ \citenamefont {Holevo}}]{giovannetti2014ultimate}%
  \BibitemOpen
  \bibfield  {author} {\bibinfo {author} {\bibfnamefont {V.}~\bibnamefont
  {Giovannetti}}, \bibinfo {author} {\bibfnamefont {R.}~\bibnamefont
  {Garc{\'\i}a-Patr{\'o}n}}, \bibinfo {author} {\bibfnamefont {N.}~\bibnamefont
  {Cerf}}, \ and\ \bibinfo {author} {\bibfnamefont {A.}~\bibnamefont
  {Holevo}},\ }\href@noop {} {\bibfield  {journal} {\bibinfo  {journal} {Nature
  Photonics}\ }\textbf {\bibinfo {volume} {8}},\ \bibinfo {pages} {796}
  (\bibinfo {year} {2014})}\BibitemShut {NoStop}%
\bibitem [{\citenamefont {Mari}\ \emph {et~al.}(2014)\citenamefont {Mari},
  \citenamefont {Giovannetti},\ and\ \citenamefont {Holevo}}]{mari2014quantum}%
  \BibitemOpen
  \bibfield  {author} {\bibinfo {author} {\bibfnamefont {A.}~\bibnamefont
  {Mari}}, \bibinfo {author} {\bibfnamefont {V.}~\bibnamefont {Giovannetti}}, \
  and\ \bibinfo {author} {\bibfnamefont {A.~S.}\ \bibnamefont {Holevo}},\
  }\href@noop {} {\bibfield  {journal} {\bibinfo  {journal} {Nature
  communications}\ }\textbf {\bibinfo {volume} {5}} (\bibinfo {year}
  {2014})}\BibitemShut {NoStop}%
\bibitem [{\citenamefont {Giovannetti}\ \emph {et~al.}(2015)\citenamefont
  {Giovannetti}, \citenamefont {Holevo},\ and\ \citenamefont
  {Mari}}]{giovannetti2015majorization}%
  \BibitemOpen
  \bibfield  {author} {\bibinfo {author} {\bibfnamefont {V.}~\bibnamefont
  {Giovannetti}}, \bibinfo {author} {\bibfnamefont {A.~S.}\ \bibnamefont
  {Holevo}}, \ and\ \bibinfo {author} {\bibfnamefont {A.}~\bibnamefont
  {Mari}},\ }\href@noop {} {\bibfield  {journal} {\bibinfo  {journal}
  {Theoretical and Mathematical Physics}\ }\textbf {\bibinfo {volume} {182}},\
  \bibinfo {pages} {284} (\bibinfo {year} {2015})}\BibitemShut {NoStop}%
\bibitem [{\citenamefont {Holevo}(2015)}]{holevo2015gaussian}%
  \BibitemOpen
  \bibfield  {author} {\bibinfo {author} {\bibfnamefont {A.~S.}\ \bibnamefont
  {Holevo}},\ }\href@noop {} {\bibfield  {journal} {\bibinfo  {journal}
  {Uspekhi Matematicheskikh Nauk}\ }\textbf {\bibinfo {volume} {70}},\ \bibinfo
  {pages} {141} (\bibinfo {year} {2015})}\BibitemShut {NoStop}%
\bibitem [{\citenamefont {De~Palma}\ \emph
  {et~al.}(2016{\natexlab{a}})\citenamefont {De~Palma}, \citenamefont
  {Trevisan},\ and\ \citenamefont {Giovannetti}}]{de2015passive}%
  \BibitemOpen
  \bibfield  {author} {\bibinfo {author} {\bibfnamefont {G.}~\bibnamefont
  {De~Palma}}, \bibinfo {author} {\bibfnamefont {D.}~\bibnamefont {Trevisan}},
  \ and\ \bibinfo {author} {\bibfnamefont {V.}~\bibnamefont {Giovannetti}},\
  }\href {\doibase 10.1109/TIT.2016.2547426} {\bibfield  {journal} {\bibinfo
  {journal} {IEEE Transactions on Information Theory}\ }\textbf {\bibinfo
  {volume} {62}},\ \bibinfo {pages} {2895} (\bibinfo {year}
  {2016}{\natexlab{a}})}\BibitemShut {NoStop}%
\bibitem [{\citenamefont {Jabbour}\ \emph {et~al.}(2015)\citenamefont
  {Jabbour}, \citenamefont {Garc{\'\i}a-Patr{\'o}n},\ and\ \citenamefont
  {Cerf}}]{jabbour2015majorization}%
  \BibitemOpen
  \bibfield  {author} {\bibinfo {author} {\bibfnamefont {M.~G.}\ \bibnamefont
  {Jabbour}}, \bibinfo {author} {\bibfnamefont {R.}~\bibnamefont
  {Garc{\'\i}a-Patr{\'o}n}}, \ and\ \bibinfo {author} {\bibfnamefont {N.~J.}\
  \bibnamefont {Cerf}},\ }\href@noop {} {\bibfield  {journal} {\bibinfo
  {journal} {arXiv preprint arXiv:1512.08225}\ } (\bibinfo {year}
  {2015})}\BibitemShut {NoStop}%
\bibitem [{\citenamefont {De~Palma}\ \emph
  {et~al.}(2016{\natexlab{b}})\citenamefont {De~Palma}, \citenamefont
  {Trevisan},\ and\ \citenamefont {Giovannetti}}]{de2016gaussian}%
  \BibitemOpen
  \bibfield  {author} {\bibinfo {author} {\bibfnamefont {G.}~\bibnamefont
  {De~Palma}}, \bibinfo {author} {\bibfnamefont {D.}~\bibnamefont {Trevisan}},
  \ and\ \bibinfo {author} {\bibfnamefont {V.}~\bibnamefont {Giovannetti}},\
  }\href {http://arxiv.org/abs/1605.00441} {\bibfield  {journal} {\bibinfo
  {journal} {arXiv preprint arXiv:1605.00441}\ } (\bibinfo {year}
  {2016}{\natexlab{b}})}\BibitemShut {NoStop}%
\bibitem [{\citenamefont {Konig}\ and\ \citenamefont
  {Smith}(2014)}]{konig2014entropy}%
  \BibitemOpen
  \bibfield  {author} {\bibinfo {author} {\bibfnamefont {R.}~\bibnamefont
  {Konig}}\ and\ \bibinfo {author} {\bibfnamefont {G.}~\bibnamefont {Smith}},\
  }\href@noop {} {\bibfield  {journal} {\bibinfo  {journal} {Information
  Theory, IEEE Transactions on}\ }\textbf {\bibinfo {volume} {60}},\ \bibinfo
  {pages} {1536} (\bibinfo {year} {2014})}\BibitemShut {NoStop}%
\bibitem [{\citenamefont {De~Palma}\ \emph {et~al.}(2014)\citenamefont
  {De~Palma}, \citenamefont {Mari},\ and\ \citenamefont
  {Giovannetti}}]{de2014generalization}%
  \BibitemOpen
  \bibfield  {author} {\bibinfo {author} {\bibfnamefont {G.}~\bibnamefont
  {De~Palma}}, \bibinfo {author} {\bibfnamefont {A.}~\bibnamefont {Mari}}, \
  and\ \bibinfo {author} {\bibfnamefont {V.}~\bibnamefont {Giovannetti}},\
  }\href
  {http://www.nature.com/nphoton/journal/v8/n12/full/nphoton.2014.252.html}
  {\bibfield  {journal} {\bibinfo  {journal} {Nature Photonics}\ }\textbf
  {\bibinfo {volume} {8}},\ \bibinfo {pages} {958} (\bibinfo {year}
  {2014})}\BibitemShut {NoStop}%
\bibitem [{\citenamefont {De~Palma}\ \emph {et~al.}(2015)\citenamefont
  {De~Palma}, \citenamefont {Mari}, \citenamefont {Lloyd},\ and\ \citenamefont
  {Giovannetti}}]{de2015multimode}%
  \BibitemOpen
  \bibfield  {author} {\bibinfo {author} {\bibfnamefont {G.}~\bibnamefont
  {De~Palma}}, \bibinfo {author} {\bibfnamefont {A.}~\bibnamefont {Mari}},
  \bibinfo {author} {\bibfnamefont {S.}~\bibnamefont {Lloyd}}, \ and\ \bibinfo
  {author} {\bibfnamefont {V.}~\bibnamefont {Giovannetti}},\ }\href
  {http://journals.aps.org/pra/abstract/10.1103/PhysRevA.91.032320} {\bibfield
  {journal} {\bibinfo  {journal} {Physical Review A}\ }\textbf {\bibinfo
  {volume} {91}},\ \bibinfo {pages} {032320} (\bibinfo {year}
  {2015})}\BibitemShut {NoStop}%
\bibitem [{\citenamefont {Audenaert}\ \emph {et~al.}(2015)\citenamefont
  {Audenaert}, \citenamefont {Datta},\ and\ \citenamefont
  {Ozols}}]{audenaert2015entropy}%
  \BibitemOpen
  \bibfield  {author} {\bibinfo {author} {\bibfnamefont {K.}~\bibnamefont
  {Audenaert}}, \bibinfo {author} {\bibfnamefont {N.}~\bibnamefont {Datta}}, \
  and\ \bibinfo {author} {\bibfnamefont {M.}~\bibnamefont {Ozols}},\
  }\href@noop {} {\bibfield  {journal} {\bibinfo  {journal} {arXiv preprint
  arXiv:1503.04213}\ } (\bibinfo {year} {2015})}\BibitemShut {NoStop}%
\bibitem [{\citenamefont {Schaller}(2014)}]{schaller2014open}%
  \BibitemOpen
  \bibfield  {author} {\bibinfo {author} {\bibfnamefont {G.}~\bibnamefont
  {Schaller}},\ }\href@noop {} {\emph {\bibinfo {title} {Open Quantum Systems
  Far from Equilibrium}}}\ (\bibinfo  {publisher} {Springer My Copy UK},\
  \bibinfo {year} {2014})\BibitemShut {NoStop}%
\bibitem [{\citenamefont {Breuer}\ and\ \citenamefont
  {Petruccione}(2007)}]{breuer2007theory}%
  \BibitemOpen
  \bibfield  {author} {\bibinfo {author} {\bibfnamefont {H.}~\bibnamefont
  {Breuer}}\ and\ \bibinfo {author} {\bibfnamefont {F.}~\bibnamefont
  {Petruccione}},\ }\href@noop {} {\emph {\bibinfo {title} {The Theory of Open
  Quantum Systems}}}\ (\bibinfo  {publisher} {OUP Oxford},\ \bibinfo {year}
  {2007})\BibitemShut {NoStop}%
\bibitem [{\citenamefont {Gogolin}\ and\ \citenamefont
  {Eisert}(2016)}]{gogolin2015equilibration}%
  \BibitemOpen
  \bibfield  {author} {\bibinfo {author} {\bibfnamefont {C.}~\bibnamefont
  {Gogolin}}\ and\ \bibinfo {author} {\bibfnamefont {J.}~\bibnamefont
  {Eisert}},\ }\href@noop {} {\bibfield  {journal} {\bibinfo  {journal}
  {Reports on Progress in Physics}\ }\textbf {\bibinfo {volume} {79}},\
  \bibinfo {pages} {056001} (\bibinfo {year} {2016})}\BibitemShut {NoStop}%
\bibitem [{\citenamefont {Wehrl}(1978)}]{wehrl1978general}%
  \BibitemOpen
  \bibfield  {author} {\bibinfo {author} {\bibfnamefont {A.}~\bibnamefont
  {Wehrl}},\ }\href@noop {} {\bibfield  {journal} {\bibinfo  {journal} {Reviews
  of Modern Physics}\ }\textbf {\bibinfo {volume} {50}},\ \bibinfo {pages}
  {221} (\bibinfo {year} {1978})}\BibitemShut {NoStop}%
\bibitem [{\citenamefont {Holevo}(2013)}]{holevo2013quantum}%
  \BibitemOpen
  \bibfield  {author} {\bibinfo {author} {\bibfnamefont {A.~S.}\ \bibnamefont
  {Holevo}},\ }\href@noop {} {\emph {\bibinfo {title} {Quantum Systems,
  Channels, Information: A Mathematical Introduction}}},\ De Gruyter Studies in
  Mathematical Physics\ (\bibinfo  {publisher} {De Gruyter},\ \bibinfo {year}
  {2013})\BibitemShut {NoStop}%
\bibitem [{\citenamefont {Ferraro}\ \emph {et~al.}(2005)\citenamefont
  {Ferraro}, \citenamefont {Olivares},\ and\ \citenamefont
  {Paris}}]{ferraro2005gaussian}%
  \BibitemOpen
  \bibfield  {author} {\bibinfo {author} {\bibfnamefont {A.}~\bibnamefont
  {Ferraro}}, \bibinfo {author} {\bibfnamefont {S.}~\bibnamefont {Olivares}}, \
  and\ \bibinfo {author} {\bibfnamefont {M.~G.}\ \bibnamefont {Paris}},\
  }\href@noop {} {\bibfield  {journal} {\bibinfo  {journal} {arXiv preprint
  quant-ph/0503237}\ } (\bibinfo {year} {2005})}\BibitemShut {NoStop}%
\bibitem [{\citenamefont {Cox}\ \emph {et~al.}(2015)\citenamefont {Cox},
  \citenamefont {Little},\ and\ \citenamefont {O'Shea}}]{cox2015ideals}%
  \BibitemOpen
  \bibfield  {author} {\bibinfo {author} {\bibfnamefont {D.}~\bibnamefont
  {Cox}}, \bibinfo {author} {\bibfnamefont {J.}~\bibnamefont {Little}}, \ and\
  \bibinfo {author} {\bibfnamefont {D.}~\bibnamefont {O'Shea}},\ }\href@noop {}
  {\emph {\bibinfo {title} {Ideals, Varieties, and Algorithms: An Introduction
  to Computational Algebraic Geometry and Commutative Algebra}}},\
  Undergraduate Texts in Mathematics\ (\bibinfo  {publisher} {Springer
  International Publishing},\ \bibinfo {year} {2015})\BibitemShut {NoStop}%
\bibitem [{\citenamefont {Bhatia}(2013)}]{bhatia2013matrix}%
  \BibitemOpen
  \bibfield  {author} {\bibinfo {author} {\bibfnamefont {R.}~\bibnamefont
  {Bhatia}},\ }\href@noop {} {\emph {\bibinfo {title} {Matrix Analysis}}},\
  Graduate Texts in Mathematics\ (\bibinfo  {publisher} {Springer New York},\
  \bibinfo {year} {2013})\BibitemShut {NoStop}%
\bibitem [{\citenamefont {Horn}\ and\ \citenamefont
  {Johnson}(2012)}]{horn2012matrix}%
  \BibitemOpen
  \bibfield  {author} {\bibinfo {author} {\bibfnamefont {R.}~\bibnamefont
  {Horn}}\ and\ \bibinfo {author} {\bibfnamefont {C.}~\bibnamefont {Johnson}},\
  }\href@noop {} {\emph {\bibinfo {title} {Matrix Analysis}}},\ Matrix
  Analysis\ (\bibinfo  {publisher} {Cambridge University Press},\ \bibinfo
  {year} {2012})\BibitemShut {NoStop}%
\bibitem [{\citenamefont {Fan}(1951)}]{fan1951maximum}%
  \BibitemOpen
  \bibfield  {author} {\bibinfo {author} {\bibfnamefont {K.}~\bibnamefont
  {Fan}},\ }\href@noop {} {\bibfield  {journal} {\bibinfo  {journal}
  {Proceedings of the National Academy of Sciences of the United States of
  America}\ }\textbf {\bibinfo {volume} {37}},\ \bibinfo {pages} {760}
  (\bibinfo {year} {1951})}\BibitemShut {NoStop}%
\end{thebibliography}%
\bibliographystyle{apsrev4-1}
\end{document}